\documentclass[useAMS,usenatbib]{mn2e}
\usepackage{psfig,latexsym,longtable,lscape}

\def\apj{ApJ}
\def\mnras{MNRAS}

\def\>{$>$}
\def\<{$<$}
\def\sun{$_{\odot}$}

\def\ergs{$\rm {erg\,s}^{-1}$}

\def\newline{\hfil\break}
\def\mincir{\ \raise -2.truept\hbox{\rlap{\hbox{$\sim$}}\raise5.truept  
\hbox{$<$}\ }}                        
\def\magcir{\ \raise -2.truept\hbox{\rlap{\hbox{$\sim$}}\raise5.truept  
\hbox{$>$}\ }}                        

\title[The Cartwheel HLX]{On the compact nature of the most luminous ULX in the Cartwheel ring}
\author[A. Wolter, G. Trinchieri and M. Colpi]{Anna Wolter$^{1}$\thanks{E-mail:
anna@brera.mi.astro.it}, Ginevra Trinchieri$^{1}$\thanks{E-mail:
ginevra.trinchieri@brera.inaf.it} and Monica Colpi$^{2}$\thanks{E-mail:
Monica.Colpi@mib.infn.it }\\
$^{1}$INAF, Osservatorio Astronomico di Brera, via Brera 28, 20121 Milano, Italy\\
$^{2}$Dipartimento di Fisica G. Occhialini, 
Universit\`a degli Studi di Milano Bicocca, Piazza della Scienza 3,
20126 Milano, Italy}
\begin{document}

\date{}

\pagerange{\pageref{firstpage}--\pageref{lastpage}} \pubyear{2006}

\maketitle

\label{firstpage}

\begin{abstract}
We report the first 
detection of flux variability in the
most luminous X-ray source 
in the southern ring of the Cartwheel galaxy.
{\it XMM--Newton} data show that the luminosity has 
varied over a timescale of six months 
from $L_{0.5-10 \rm {keV}} \sim 1.3 \times 10^{41}$ \ergs, 
consistent with the previous {\it Chandra} observation, to 
$L_{0.5-10 \rm{keV}}\mincir 6.4 \times 10^{40}$ \ergs.
This fact provides the first evidence that the source is 
compact in nature and is not a collection
of individual fainter sources, such as supernova remnants. 
The source has been repeatedly observed at the 
very high luminosity level 
of $L_{0.5-10 \rm {keV}} 
\sim 1.3 \times 10^{41}$ \ergs\, for a period of at least 4 years 
before dimming at
the current level.  
It represents then the first example of an accreting object
revealed in a long lived state of extremely high luminosity.

\end{abstract}


\section{Introduction}

Very luminous off-nuclear X-ray sources were
discovered in nearby galaxies with the Einstein satellite 
(Fabbiano et al. 1989). They were named Ultra-Luminous X-ray sources
(ULXs) based on their 
X-ray luminosities, much higher than the Eddington limit
for a solar mass black hole ($L_X \sim  1.4 \times 10^{38}$ \ergs\,).
These luminosities may reflect beamed emission from an accreting stellar 
mass compact object, or super-Eddington emission, or isotropic 
accretion onto an intermediate mass black hole.
The brightest objects, those with $L_X \magcir 10^{41}$ \ergs,
sometimes termed 
Hyperluminous X-ray sources (HLXs; 
see Matsumoto et al. 2001; Kaaret et al. 2001), are even more intriguing,
since their luminosities are closer to that of active galactic nuclei,
requiring a bigger engine, stronger beaming or even
extreme super-Eddington regimes.
The issue of the physical interpretation of ULXs and HLXs 
is still quite open.

An extraordinary example of HLX is the source N.10 detected
in the narrow, gas-rich star--forming ring of the Cartwheel galaxy
with isotropic luminosity of $L_{0.5-10 \rm{keV}} \sim  
1.3 \times 10^{41}$ \ergs\, (Wolter et al. 1999; 
Wolter \& Trinchieri 2004 - hereafter WT04; Gao et al. 2003). 
This is the brightest of a number of individual sources that 
also appear to reside in the ring, all classified as
ULXs, based on their isotropic luminosities in excess 
of $L_{0.5-10{\rm keV}} = 3 \times  10^{39}$ \ergs\, (WT04).

The spatial association of the N.10 HLX and the 
ring--ULXs with HII complexes and  
young star-forming clusters suggests a physical link, thus
providing an invaluable in-depth probe of the
young stellar population currently present in the Cartwheel outer ring.
The physical nature of the ULXs in the Cartwheel is not clear 
yet: they could be genuine single sources, hence
accreting compact objects, or unresolved collections of supernova
remnants.  A way to disentangle this puzzle is to look at their
variability.  Variability in ULXs, covering timescales of months to a
few years, has been often encountered (e.g., in the Antennae; Fabbiano
et al. 2003) and was taken as evidence that they represent an
apparently bright state of accreting high mass X-ray binaries (HMXBs)
caught in some peculiar evolutionary stage (King 2002).  Young, short
lived, HMXBs hosting a neutron star or a stellar-mass black hole
are common in star--burst galaxies and their number is likely to be
linked with the star--formation rate 
of their hosting galaxy (Fabbiano 2005). 

It has been proposed that HMXBs share a
universal X-ray luminosity function that extends up to luminosities of
$10^{40}$ \ergs, characteristic of ULXs (Grimm, Gilfanov, \& Sunyaev
2003).
WT04 have
shown that the X-ray sources seen in the Cartwheel follow the
same luminosity function 
and derive a star formation rate of 20 M\sun yr$^{-1}$ 
in agreement with recent radio and Far Infrared estimates (Mayya et al. 2005).
The only outlier 
appears to be source N.10, brighter than expected at the 
bright end of the X-ray Luminosity Function (XLF) by a factor of 3.
WT04 noticed that a higher cutoff would account 
for this excess in the Cartwheel XLF,
which would otherwise highlight a different
physical origin for this exceptional HLX.

At present, the only confirmed HLX reported in the literature is the 
source X--1 
in the starburst galaxy M82, with
isotropic luminosity of  
a few $\times 10^{40}$ \ergs\, (Ptak \& Griffiths 1999: ASCA baseline flux;
Strohmayer \& Mushotzky 2003; Dewangan et al. 2006: 
{\it XMM--Newton} observations).
X--1 in M82 showed a burst, up to $10^{41}$ \ergs, lasting about
a month during ASCA observations (Ptak \& Griffiths 1999)
and again with {\it Chandra} (Matsumoto et al. 2001; Kaaret et al. 2001).
Spectral and temporal variability points to a 200M\sun\, intermediate 
mass black hole as the accreting object (Dewangan et al. 2006).  
If confirmed by additional observations,
X--1 is the prototype of a ``new" class of black holes,
bridging the gap between those of stellar origin and the very massive
black holes hosted in galactic nuclei (Miller \& Colbert 2004).
Other candidate HLXs are those in  
NGC 7714 ($L_X = 6.\times 10^{40}$ erg s$^{-1}$ in the high state; Soria 
\& Motch 2004; Smith, Struck \& Nowak, 2005),
in NGC2276 ($L_X = 1.3\times 10^{41}$ erg s$^{-1}$ in the 0.5-10 keV band;
Davis \& Mushotzky 2004) and in MCG-03-34-63 
(ULX 1 with peak luminosity of $L_X = 1.3\times 10^{41}$ erg s$^{-1}$ in 
the 0.5-7 keV band; Miniutti et al. 2006). 
The majority of these objects have been observed with {\it XMM--Newton}
only, and with limited time coverage; 
therefore neither a secure physical explanation nor a clear
association with the host galaxy can be put forward.

The Cartwheel source N. 10 is possibly the brightest HLX known 
but the lack of detailed information on its variability has prevented 
the exclusion of an extended nature. The new 
data that we present here have now confirmed its compact nature.

The X-ray image of the Cartwheel shows, besides a number of point sources,
the presence of hot ($k_B T$=0.2 keV)
gas in the ring with a luminosity of  
$L^{gas}_{0.5-2 {\rm keV}} = 3 \times 10^{40}$ \ergs\, (WT04) 
and suggests that more gas
might be present in the near environment of the galaxy.  To better study
all of these components we have obtained new {\it XMM--Newton} observations.  
Here we concentrate on the properties of the HLX, leaving a more
comprehensive presentation of the galaxy as a whole to a forthcoming
paper.  For the Cartwheel we use  a distance D=122 Mpc (calculated
from Amram et al. 1998, with H$_{\rm 0}$ = 75 km s$^{-1}$ Mpc$^{-1}$) 
corresponding to a linear scale of 0.592 kpc/arcsec.

\section{XMM data}

The  Cartwheel was observed with {\it XMM--Newton}  on the 14-15th 
of December, 2004 (36 ksec; dataset [101]) and on the 21st-22nd May, 2005 
(60 ksec; dataset [201]),
with the pn and MOS instruments operating 
in full-frame mode with the thin filter applied.
Standard rejection criteria were applied to eliminate high background periods
that are fortunately relatively short.
The net exposure times, 
after data cleaning, reduce to 29(24) and 50(42) ks
for MOS(pn), a $\sim$ 20\% reduction in time.
 
The two datasets were analyzed independently. We have used the 
standard XMM Science Analysis System tasks 
to prepare the data and produce images and spectra.
Our primary interest, source N.10 in the {\it Chandra} list (WT04), falls 
 over a bad  column in MOS2 in the first
observation, and in MOS1 in the second one. We will therefore consider
only MOS1 and pn for the [101] observation and MOS2 and pn for the [201] 
observation. 

We have checked consistency between the two observations by using different
regions within and outside the Cartwheel. The aspect solutions of the two
dataset are very similar, so that source positions do not differ between [101] and [201]. Bright field sources are expected
to be mostly AGN, therefore flux variations of the order of a factor of 2 
are common, so no single source can be used as a reference point.
However, we have confirmed that overall the count rates are constant
for a large number of sources between the two observations.

\subsection{Image}

 \begin{figure*}
 \scriptsize 
 \psfig{file=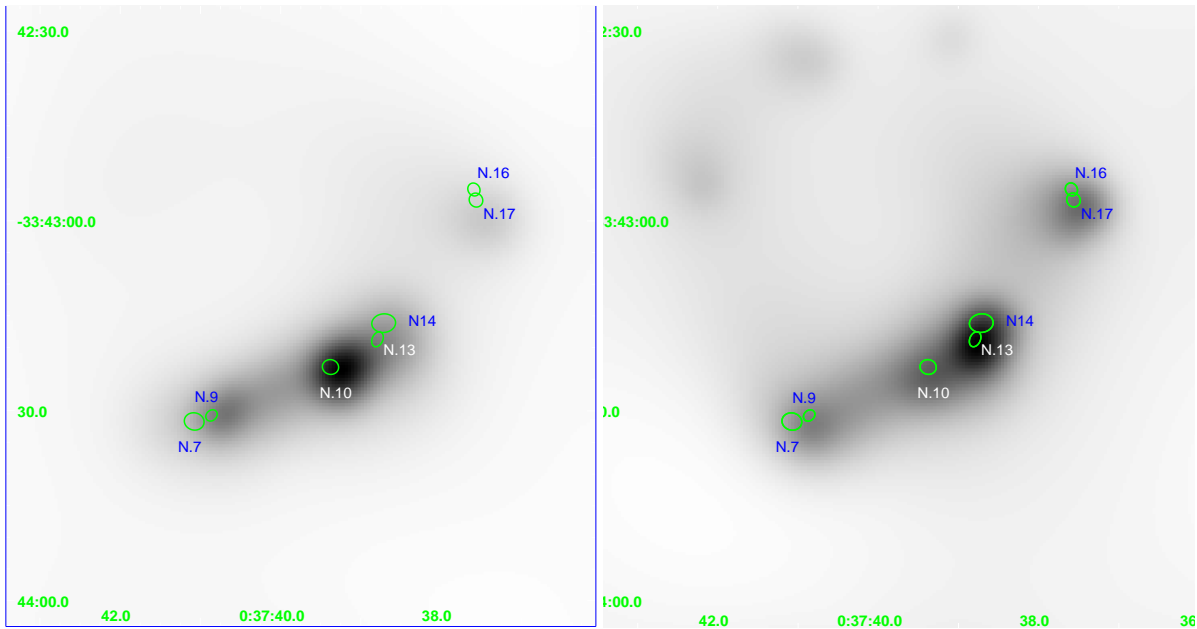,width=16cm}
 \caption{A smoothed grey scale image of pn data for the southwestern ring from the 
 first observation [101], {\bf left}, and the second [201]. {\bf right}.
 The positions of the brightest {\it Chandra} sources are indicated. No
 correction for different aspect solutions has been attempted. Distances 
 between {\it Chandra} and {\it XMM--Newton} peaks are below aspect 
uncertainties.
 In the [101] data, N.10 is clearly the brightest source; it is unresolved
by  {\it XMM--Newton}.
 In the  [201] observation, which is longer, more details and more sources
 appear. The detection of a faint source at the position of N. 10 is 
difficult due to the close vicinity of the brighter sources N. 13\&14.
 Notice also the relative intensity of sources N.7\&N.9 and 
 sources N.16\&N.17 which
 is varied between the two observations. Sources  N.7\&N.9,  N.13\&N.14,
and N.16\&N.17 cannot be separated by {\it XMM--Newton}. }
 \label{ima}
 \end{figure*}

We compare the pn images from the [101] and [201] 
observations, smoothed 
with an adaptive Gaussian kernel (package 
{\em csmooth} from Ciao 3.3). In Fig.~\ref{ima} we plot
the two images side by side. We overplot for reference the positions 
of {\it Chandra} sources, that are not at the peak of the {\it XMM--Newton}
positions. In fact, no
correction for the different aspect solutions of the two satellites
has been attempted, however 
distances between {\it Chandra} and {\it XMM--Newton} peaks are below aspect 
uncertainties .
Although the two images should not be used for 
a quantitative comparison since
the two 
observations have different lengths and should be properly 
normalized and corrected for possible background differences, the graphical 
comparison shows variability in different areas.
In particular, it is evident 
that the HLX is no longer the brightest
source in the second observation. 
From the comparison between the count rates in
the neighbouring sources, we have determined that the source next to it,
which corresponds to {\it Chandra} sources N.13 \& N.14, has not varied between
the two observations (the count rate in the second observation is at most 15\%
higher than in the first one). If we assume that this is constant, then
the source to the NW, corresponding to N.16 \& N.17, is also constant (the
same 15\% increase in the count rate), but the SE source (N.7 \& N.9) 
has faded to
about half its strength in the second observation.  Note that the darker
colours in the image simply reflect the higher statistics available in the
second exposure, which is about twice in length.

\subsection{Spectrum}

 \tabcolsep 0.3cm
 \begin{center}
 \begin{table*}
 \caption{Log of {\it XMM--Newton} Observations\label{tab1}}
 \begin{center}
 \begin{tabular}{l r l r r }
 \hline
\hline
 Name & Date & Instrument & Net Counts$^{a}$  & Exp. Time   \\
          &   &  & (0.5-10 keV)  & sec   \\
\hline
 \\              
 N.10. & 12/14/2004 &  MOS1  &  80.8$\pm$9.5    & 29,583 \\ 
       & [101]    &  pn      & 202.1$\pm$14.6   & 24,418  \\
 N.10. & 05/21/2005 & MOS2   &  87.2$\pm$9.9    & 49,669  \\
       & [201]    &  pn      & 237.9$\pm$15.9   & 42,277   \\
\hline
 \end{tabular}

\smallskip
 $^{a}$ {in a 10$^{\prime\prime}$ radius, centred on RA(2000) = $00^{h}\, 37^{m}\, 39^{s}\!\!.38$
and Dec(2000) = $-33^{\circ}\, 43^{\prime}\, 23\farcs08$, see text.}
 \end{center}
 \end{table*}
 \end{center}

We extract spectra from a region of radius 10$^{\prime\prime}$ centred 
about the peak of N.10 in [101] and background from a 
nearby circular region devoid of sources. 
Appropriate response matrices for spectral analysis
where generated using the SAS tasks  {\it arfgen} and {\it rmfgen}. 
To improve on the statistics, we have binned the data so that each bin
has a significance of at least 2$\sigma$.
We report in Table~\ref{tab1} the total net counts and exposure times 
in seconds, after cleaning, for this extraction region.

The extraction radii for {\it XMM--Newton} 
are smaller than customary, due to the presence 
of many surrounding sources in the Cartwheel ring.
Nonetheless they include a portion of the ring,
so we expect a fraction
of the diffuse underlying gas component to contaminate the HLX spectrum.

We first fit the [101] spectrum with a simple model, i.e. an absorbed
power law. This results in an acceptable fit ($\chi^{2} = 15.7$ for 23 d.o.f)
with a slope $\Gamma = 1.75 \pm 0.25$, low energy
absorption due to an intervening column with N$_{\rm H} = 1.4 \times 10^{21}$
cm$^{-2}$ and flux f$_{0.5-10} = 6.9 \times 10^{-14}$ erg 
cm$^{-2}$ s$^{-1}$. 
This is consistent with the {\it Chandra} results for source N.10 suggesting that
the HLX is the dominant source of emission.
However, due to the larger area considered  and larger PSF, we
expect some contribution from the underlying ring emission.
As will
become clearer later, this assumption is also relevant for  the
comparison between the two {\it XMM--Newton} observations. We have therefore
considered a spectrum that includes all three components derived in the
{\it Chandra} data: 1) gas, 2) unresolved binary sources,3) HLX (from WT04).

 \tabcolsep 0.3cm
 \begin{center}
 \begin{table}
 \caption{Fluxes of different fit components}
 \label{tab2}
 \begin{center}
 \begin{tabular}{l r r }
 \hline
 \hline
 Component  & Flux & Flux   \\
             &(0.5-2 keV)  & (2-10 keV)   \\
 \hline
 HLX [101]    & $2.4 \times 10^{-14}$  & $ 5.0 \times 10^{-14}$   \\ 
 HLX [201]    & $1.2 \times 10^{-14}$  & $ 2.3 \times 10^{-14}$   \\
 gas    & $2 \times 10^{-15}$ & --     \\
 unresolved binaries & $1 \times 10^{-15}$  & $1.3 \times 10^{-15}$    \\
 \hline  
 \end{tabular}
 \end{center}
 \end{table}
 \end{center}

Given the limited quality of the data, we have fixed 
the parameters of components 1) and 2) 
to the same values of WT04.
For the relative  normalizations we used
the {\it Chandra} values, properly rescaled to the area covered by the
present region (1/10 of 
the emission detected by {\it Chandra} in the whole ring).
The HLX normalization is left free, while $\Gamma$ is fixed at the {\it 
Chandra} value. The details of the model are shown graphically in the 
rightmost panel in Fig. 2.

The HLX is indeed the brightest component in the 10$^{\prime\prime}$ radius
region; see Table~\ref{tab2} for fluxes of the different components.
The intrinsic luminosity of the HLX is $L_{0.5-2 \rm {keV}} = 4.6 \times 10^{40}$\ergs\,
and  $L_{2-10 \rm {keV}} = 8.7 \times 10^{40}$ \ergs.

 \begin{figure*}
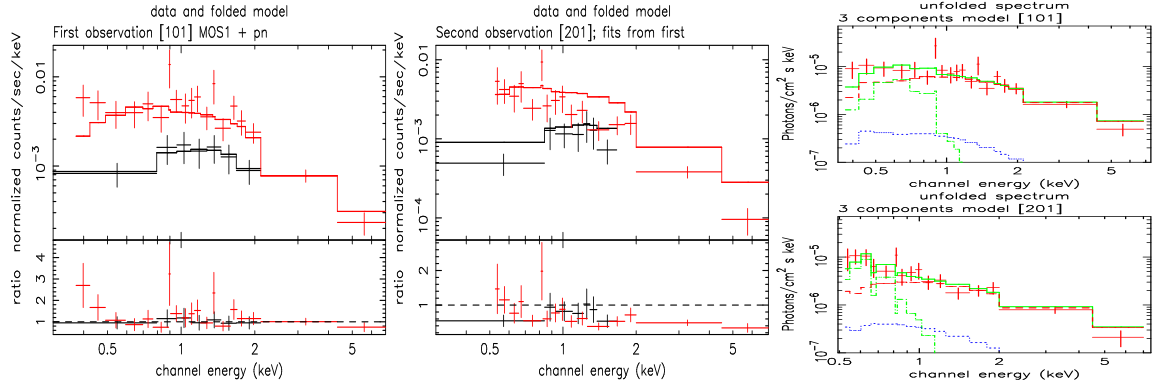

 \scriptsize 
 \hbox{
 \psfig{file=f2a.eps,width=5cm,height=5cm,angle=-90}
 \psfig{file=f2b.eps,width=5cm,height=5cm,angle=-90}
 \vbox{
 \psfig{file=f2c.eps,width=5cm,height=2.5cm,angle=-90}
 \psfig{file=f2d.eps,width=5cm,height=2.5cm,angle=-90}
 \vfill}
 \hfill}
 \caption{Left: spectrum of [101], fit as described in the text.
 Centre: spectrum of [201] with fit from [101]. Right: Unfolded spectrum,
from pn data only for clarity, showing the three components at the best
fit values. The difference between top [101]
 and bottom [201] is only the normalization of the `HLX' component [and the
binning scheme]. }
 \label{sp}
 \vskip 0.8cm
 \end{figure*}

We then used the
same extraction region, and the same binning scheme for the [201] dataset.
We obtained about the same number of net counts in spite of the longer
observing time. Again, the single power law gives an acceptable fit
($\chi^2 =18.8$ for 21 d.o.f.) with parameters consistent to [101],
and a flux about half.
If we apply the same complex spectral model of [101], with the same 
parameters, we find a large discrepancy with the data 
($\chi^2 > 170$ for 21 d.o.f.) as shown in Fig.~\ref{sp} (Centre).  
In particular we notice that the points above 0.8 keV are
systematically lower than the model.  If we make the reasonable assumption
that both the diffuse gas and the unresolved point source components 
have not varied, and let only the normalization for the HLX component vary, we
obtain a good fit for a normalization that is about 1/2 that of [101]
($\chi^2$ = 21.13 for 23 d.o.f.).
Given the data quality and the complexity of the spectral model, testing a
spectral variation in this component is unrealistic.  We therefore cannot
comment on spectral variations between a high and a low state expected for
binary sources.  However we can reasonably state that the HLX has dimmed
by at least a factor of two in the 6 months between the two observations.
The unabsorbed flux of N.10 in this second observations at the formal best 
fit values is reported in Table~\ref{tab2}.
The corresponding luminosity 
is at most $L_{0.5-2 \rm {keV}} = 2.0 \times 10^{40}$ 
\ergs\, and  $L_{2-10 \rm {keV}} = 3.8 \times 10^{40}$ \ergs.

\medskip

\subsection{Light curve}

\begin{figure}
 \scriptsize 
 \psfig{file=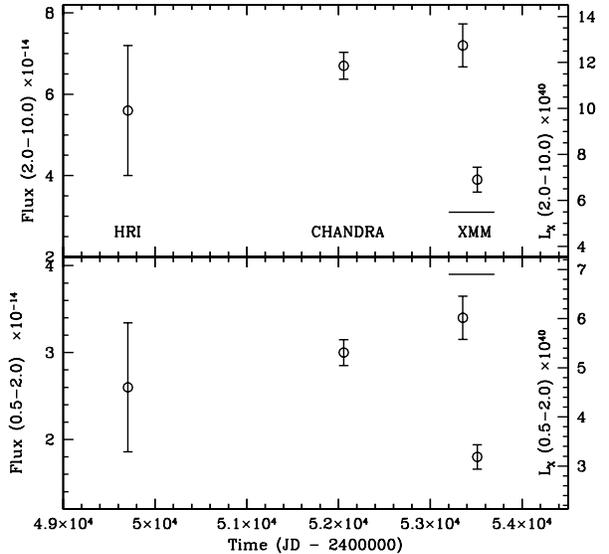,width=8cm,height=8cm}
 \caption{Long term light curve in the soft (lower panel) and hard
 (upper panel) energy
 band, over an interval of about 10 years. The two {\it XMM--Newton} points that 
 define the variation are
 not subject to cross-calibration uncertainties. All fluxes are computed 
 with the same spectral shape,
 i.e. power law with $\Gamma = 1.6$ and N$_{\rm H} = 3.6 \times 10^{21}$ 
 cm$^{-2}$; see fit to {\it Chandra} data in WT04. Right axis reports 
luminosities computed assuming the Cartwheel distance.
 }
 \label{lc}
 \end{figure}

 \tabcolsep 0.3cm
 \begin{center}
 \begin{table*}
 \caption{Summary of all observations of N.10}
 \label{Tab_lc}
 \begin{center}
 \begin{tabular}{l r l c c c}
 \hline
 \hline
 Date & JD & Instrument & Count rate  & Flux  & Flux \\
        &   &   & (0.3-2.5 keV)  & (0.5-2 keV) & (2-10 keV)  \\
 \hline
 \\              
 1994 Dec 9-23 & 2449696-2449710 & {\it ROSAT} HRI & $3.79 \pm 1.08 \times 10^{-4}$ &$ 2.6 \times 10^{-14}$ & $5.6 \times 10^{-14}$  \\
 2001 May 26-27 & 2452056-2452057 & {\it Chandra} ACIS-S & $5.30 \pm 0.26 \times 10^{-3}$  & $3.0 \times 10^{-14}$ & $6.7 \times 10^{-14}$\\
 2004 Dec 14-15 & 2453354-2453355&  {\it XMM--Newton} pn & $7.8 \pm 0.57 \times 10^{-3}$ & $3.4 \times 10^{-14}$ & $7.2 \times 10^{-14}$\\ 
 2005 May 21-22 & 2453512-2453513 &  {\it XMM--Newton} pn & $4.2 \pm 0.33 \times 10^{-3}$ &$1.8 \times 10^{-14}$ & $3.9 \times 10^{-14}$\\ 
 \hline   \\
 \end{tabular}
 \end{center}
 \end{table*}
 \end{center}

The evidence for variability during the {\it XMM--Newton} observation
prompted us to look at the long term behavior of the source.
We construct the light curve of source N.10  by using all the available 
datasets (Fig.~\ref{lc}). We extract net counts from a 10$^{\prime\prime}$ radius region
in the 0.3-2.5 keV band, to match the limited energy band of
the HRI,  
and use a nearby large circle devoid of
sources for background. We convert count rates to flux by using the
{\it Chandra} fit with a power law and conversion factor from PIMMS 
(see WT04). 
Of course cross calibration uncertainties are possible between
different instruments, and we cannot be sure that the shape of the emission
was the same at all times, however these uncertainties are probably
small in the band considered. 

Dates, instruments, count rates and fluxes in both the soft and hard
band, extrapolated from the same model, are reported in Tab.~\ref{Tab_lc}.
From inspection of the light curve we deduce that the HLX was in a
brighter state from 
2001 to 2004. The HRI errorbar
is consistent at the lowest level with a ``ring-only'' contribution,
but an enhancement of the emission was visible in the area in the HRI
data, to indicate that the source was probably also in this bright state
since 1994
(see Wolter et al. 1999). 
The source became significantly fainter in the six months between the 
2004 and 2005
{\it XMM--Newton} observation, assuming a ``ULX'' status.

\medskip

\section{Discussion}

Spectral properties and variability in some ULXs 
(e.g. Makishima et al. 2000) 
suggest that we are witnessing
accretion onto a compact object, in a binary system.
However, no universal model
exists. ULXs might be stellar mass black holes with anisotropic X-ray emission
due to mechanical beaming
(King et al. 2001), or relativistic beaming from a jet 
(Mirabel \& Rodriguez 1999;  K\"ording
et al. 2002), or  stellar mass black holes accreting at super-Eddington 
rates (Begelman 2002).
The most challenging model for the HLXs, given their extreme
luminosities, is that of a binary
system hosting a 10$^{2-4}$ M\sun\, 
black hole (e.g. Colbert \& Mushotzky 1999).
Intermediate mass black holes may form from the collapse
of very massive stars born through
stellar runaway collisions in
dense star clusters (Portegies Zwart \& McMillan 2002;
Gurkan et al. 2004). In such a young environment, 
the intermediate mass black hole may gain a
massive donor star ($\magcir 20$ M\sun) 
through tidal events (Baumgardt et al. 2005)
or dynamical interactions (Blecha et al. 2006).
This system will be able to
sustain luminosities as high as $10^{40-41}$ \ergs\, as seen in the 
very bright ULXs (Patruno et al. 2005; Madhusudhan et al. 2006). 
Spatial association of X-ray sources with young star cluster has been
searched for a number of ULXs (Kaaret et al., 2004) 
but has not been firmly established yet.
The whole ring of the
Cartwheel consists of bubbles and condensations (Struck et al. 1996) and
the neighbourhood of N.10 is no exception. Given the distance of the 
galaxy any small misalignment between X-ray and optical (HST) positions
implies kpc scale distances, so a precise determination of the optical
counterpart is hard, without a proper absolute cross-calibration
of the two images. 
In any case, the association with an environment of massive and young
stars is almost certain.

The 
{\it XMM--Newton} observatory has revealed a factor of two dimming in the
flux of source N.10 in the Cartwheel, over a
timescale of $\sim 6$ months. Although this kind of variability
has been observed in other ULXs, in the N.10 case it provides the first
evidence against the hypothesis
that the source is a chance superposition of fainter sources such as
young supernova remnants, and suggests that it is a truly compact 
source. 
The estimated age of the southern ring
($<10$ Myr), 
and the lack of radial spread of its sources, indicate that source N.10 
is closely linked to the active star forming episode and its youth 
suggests that the hypothetical intermediate mass black hole  
present in N.10   
had a high chance of capturing the massive companion star through a tidal
event (Baumgardt et al. 2005).
The decline in flux by a factor of two brings
the source down to the luminosity level of many other ULXs.
So, one may suspect that variability 
(which is seen in the directions of both increasing or decreasing
luminosity; Fabbiano et al. 2003) 
occasionally transform ULXs in HLXs and viceversa.
The dimming of source N.10 is such that its flux is now consistent with
the star--formation--rate normalized  XLF
of the Cartwheel (assumed to be constant as observed in other
well monitored sources such as the Antennae, Zezas et al. 2004), demonstrating continuity,  
at the brightest fluxes, with the HMXBs and ULXs. 
Most probably, then, ULXs and HLXs are low and high states respectively of the
same class of sources. Continuity in the XLF with HMXB suggests also that
ULXs might not be an entirely different phenomenon: they may represent 
extreme  high luminous states related to the fainter accreting
binaries.
At present we can not exclude 
that source N.10 is powered by
an intermediate mass black hole since we lack
detailed information on the spectrum and on the variability properties 
due to the limited statistics available.
Deep optical and radio observations might give
further insight into the nature and the environment of such source. 
On the other hand, 
source N. 10 may be a HMXB 
accreting anisotropically onto a stellar mass black hole in a peculiarly high
luminosity 
state (King 2002, 2004). 
We can reject instead the possibility that a rotation powered Crab-like pulsar
(Perna \& Stella 2004)  
is hosted in the HLX, since a luminosity decay by
magnetic braking over a time scale of six months would require
the occurrence of a young pulsar of comparable age.
This is inconsistent with the 
stability of the light curve observed over the last 4 or 10 years, from the
{\it ROSAT}, {\it Chandra} and {\it XMM--Newton} 
composite data.

King \& Dehnen (2005) suggested recently that
HLXs differ substantially from ULXs, and are
naked, tidally stripped nuclei of dwarf galaxies 
hosting a massive relatively bright black hole. For 
substantial tidal stripping to occur, a very close
distance of approach ($\sim 200$ pc) is required for the impinging dwarf
inside the main galaxy.
This is just the distance of X--1 from the core of M82,
which the authors interpret as an active relic of a naked  dwarf core. 
For source N.10 this scenario runs into serious difficulties.
The southern ring of the Cartwheel is a coherent expanding wave
of star formation associated to a strong gaseous density wave  
excited by a collisional perturbation with a nearby galaxy. 
Given the gas-dynamical origin of the ring it is unrealistic 
to believe in a chance coincidence of source N.10 with a stripped core
of a dwarf passing by.
We have also investigated the possibility that we are observing
a background AGN. From the extragalactic LF (Hasinger et al. 1993)
we derive a chance coincidence for a background 
source of $\leq 2\times 10^{-3}$ for the whole ring 
at the flux of the {\it Chandra} detection of source N.10. 
This is a small but non negligible possibility, which however we 
consider unlikely.

Individual most luminous ULXs, termed here HLXs, 
may represent occurrences of accretion episodes
onto intermediate mass black holes.  Source N. 10 in the Cartwheel,
together with source X-1 in M82, could be the cleanest example.
Whereas source X-1 in M82 stayed in the HLX state for timescales
of hours, source N.10 in the Cartwheel is the longest lived HLX, 
having been observed in a time frame of at least 4 years to be
as bright as $L_{0.5-10 \rm {keV}} \sim 1.3 \times 10^{41}$ \ergs. 
We thus expect that other bright ULXs may share variability
of this kind and become so bright to be observed at the HLX level.
We cannot establish at the moment whether HLXs represent an altogether 
physically distinct class of objects as suggested e.g. by King \& 
Dehnen (2005), 
or an higher luminosity transient--state of ULXs.

Repeated observations of the Cartwheel at high resolution 
are the best way to properly determine the variability 
pattern of source N.10 
and the best mean to properly study this extreme
source, which could be the first example of a long lived HLX in our
local universe.

\section*{Acknowledgments}

We thank Angela Iovino for useful discussions. 
We  acknowledge partial financial support from the Italian
Space Agency (ASI) under contract ASI-INAF I/023/05/0.
This research has made use of SAOImage DS9, developed by the Smithsonian
Astrophysical Observatory.
This work is based on observations obtained with XMM-Newton, an ESA
science mission with instruments and contributions directly funded by
ESA Member States and the USA (NASA).

\bsp

\label{lastpage}

\end{document}